\UseRawInputEncoding
\documentclass[prd,aps,preprint,amsmath,nofootinbib,amssymb,eqsecnum,showkeys,tightenlines]{revtex4-1}

\usepackage{amsmath, latexsym, amssymb, hyperref, graphicx, color, multirow, makecell, diagbox}
\usepackage[table]{xcolor}
\usepackage{tabularx}
\usepackage[T1]{fontenc}
\usepackage[capitalize]{cleveref}
\definecolor{nicered}{rgb}{.7,.1,.1}
\definecolor{nicegreen}{rgb}{.1,.5,.1}
\definecolor{darkblue}{rgb}{0,0,.5}
\hypersetup{colorlinks, citecolor=nicegreen,linkcolor=nicered, urlcolor=darkblue}
\usepackage{multirow}
\usepackage{slashed}
\numberwithin{equation}{section}
\usepackage{verbatim}
\allowdisplaybreaks

\begin{document}
\preprint{}

\title{Muon Anomalous Magnetic Moment and Higgs Potential Stability in the 331 Model from $SU(6)$}

\author{Tianjun Li$^{1,2}$}
\email{tli@mail.itp.ac.cn}

\author{Junle Pei$^{1,2}$}
\email{peijunle@mail.itp.ac.cn}

\author{Wenxing Zhang$^{3}$}
\email{zhangwenxing@sjtu.edu.cn}

\affiliation{$^1$ CAS Key Laboratory of Theoretical Physics, Institute of Theoretical Physics, Chinese Academy of Sciences, Beijing 100190, China}
\affiliation{$^2$ School of Physical Sciences, University of Chinese Academy of Sciences,
	No.~19A Yuquan Road, Beijing 100049, China}
\affiliation{$^3$ Tsung-Dao Lee Institute and School of Physics and Astronomy, Shanghai Jiao Tong University, 800 Dongchuan Road, Shanghai, 200240, China}

\begin{abstract}

We consider a $SU(3)_c \times SU(3)_L \times U(1)_X$ model from a $SU(6)$ Grand Unified Theory (GUT). In order to explain the anomalous magnetic moments of muon and electron, we introduce two new scalar triplets without vacuum expectation values (VEVs) so that the leading contributions to $\Delta a_{\mu}$ and $\Delta a_{e}$ can avoid the suppression from small muon mass. In addition, the Higgs potential stability of this 331 model is studied by giving a set of sufficient conditions to ensure the boundedness from below of the potential.

\end{abstract}

\maketitle

\section{Introduction}\label{sec:intro}

The standard model (SM) has been proved to be a successful theory since the last standard model particle, Higgs, was found at the Large Hadron Collider (LHC). 
However, there are many problems that SM cannot explain. For instance, the gauge hierarchy problem, dark matter, neutrino masses, gauge coupling unification, etc. There are so many motivations that make people believe the existence of new physics in higher energy beyond SM. In Grand Unified Theory (GUT), it is reasonable to extend the non-Abelian gauge group $SU(2)_L$ into a larger non-Abelian gauge group $SU(3)_L$ \cite{Langacker:1980js,Deppisch:2016jzl, Li:2019qxy, Huang:1994zg, Huang:1993qx,Cao:2016uur,Pleitez:1994pu,Ponce:2002sg,Dong:2006mg,Boucenna:2015zwa,PhysRevD.8.484,PhysRevD.10.1310}, where the $U(1)_Y$ is substituted by $U(1)_X$. The X charges of Higgs sector are critical in generating the breaking of $SU(3)_L$ gauge symmetry \cite{Diaz:2004fs,Fonseca:2016tbn}. Thus, it is necessary to find a natural way to obtain the representations of such particles instead of giving the representations by hand. In the 331 model proposed in  \cite{Li:2019qxy,Deppisch:2016jzl,Sen:1983xj}, all the particle representations under the $SU(3)_C \times SU(3)_L \times U(1)_X$ gauge group could be obtained from some simple representations of the $SU(6)$ group \cite{Sen:1983xj}. The extended Higgs sector brings exotic gauge bosons, in which the $Z'$ is constrained to be heavier than $\sim$ 4.5 TeV \cite{Sirunyan:2018nnz}. The exotic fermion sector also has rich new physical phenomenology, such as dark matter and neutrino masses and mixing that we have discussed in \cite{Li:2019qxy}. In this work, we will study how to explain the muon anomalous magnetic moment, $g_{\mu}-2$, in this 331 model as well as the Higgs potential stability. 

Although there is no doubt that the SM is consistent with most of the experimental data so far, the muon anomalous magnetic moment, $a_{\mu}=\frac{g_{\mu}-2}{2}$, is a long-standing deviation \cite{Tanabashi:2018oca, Aoyama:2020ynm, Grange:2015fou, Bennett:2006fi,Dutta:2020scq}. Recently, the first results from the muon g-2 experiment at the Fermi-Lab has been announced. Combining with the previous results by the Brookhaven National Lab (BNL), the measured value deviates from the theoretical prediction of SM around 4.2$\sigma$ and 
the discrepancy is given by \cite{PhysRevLett.126.141801}
\begin{equation}
\Delta a_{\mu}=a_\mu^{\text{EXP}}-a_\mu^{\text{SM}}=(2.51\pm 0.59) \times 10^{-9}.
\end{equation}

At the same time, a 2.4$\sigma$ discrepancy between the experimental and theoretical values of the electron anomalous magnetic moment, $a_e=\frac{g_e-2}{2}$, is given by \cite{Hanneke_2008, PhysRevLett.100.120801, Aoyama:2017uqe}
\begin{equation}
\Delta a_e=a_e^{\text{EXP}}-a_e^{\text{SM}}=(-8.7 \pm 3.6) \times 10^{-13}.
\end{equation}

In order to explain these deviations, models beyond-the-SM (BSM) are proposed \cite{Dutta:2020scq,
	Aoyama:2017uqe,Davier:2019can,Keshavarzi:2018mgv,Blum:2018mom, Davier:2017zfy,Fajfer:2021cxa,Jana:2020pxx,Kiritsis:2002aj,Cao:2021lmj,Das:2021zea,Chen:2021jok,Giudice:2012ms,Zhu:2021vlz,Han:2021gfu,Gu:2021mjd,Cox:2021gqq,Wang:2021bcx}. Basically, the deviations are accounted by loop diagrams involved the BSM fermions, scalars, and gauge bosons.

In the rest of the paper, we will first review this 331 model briefly in Section~\ref{331model}. The stability of Higgs potential is studied in Section~\ref{BFB}. In Section~\ref{gg2} and Section \ref{gge}, the anomalous magnetic moments of muon and electron are explained, respectively. 
Finally, we conclude in Section~\ref{conclu}.

\section{The Model}\label{331model}

As the $SU(3)_C \times SU(3)_L \times U(1)_X$ gauge group of the 331 model proposed in \cite{Li:2019qxy,Deppisch:2016jzl,Sen:1983xj} comes from a $SU(6)$ gauge group, representations of the $SU(6)$ group can be decomposed into representations of
the $SU(3)_C \times SU(3)_L \times U(1)_X$ group to make up of the fermions of this 331 model,
\begin{align}
\bar{6}&\to
(1,~\bar{3},~\frac{-1}{2\sqrt{3}})~\bigoplus~(\bar{3},~1,~\frac{1}{2\sqrt{3}})\\
&\hookrightarrow f_i=\left(e_{Li},-\nu_{Li},~N_{i}\right)~\bigoplus~ d_{Ri}^c~,\\
\nonumber\\
\bar{6}^\prime&\to(1,~\bar{3},~\frac{-1}{2\sqrt{3}}) ~\bigoplus~ (\bar{3},~1,~\frac{1}{2\sqrt{3}})\\
&\hookrightarrow f_i^\prime=\left(e_{Li}^\prime,-\nu_{Li}^\prime,~N_{i}^\prime\right) ~\bigoplus~ D_{Ri}^c~,\\
\nonumber\\
15&\to (3,~3,~0)~\bigoplus~(1,~\bar{3},~\frac{1}{\sqrt{3}})~\bigoplus~(\bar{3},~1,~\frac{-1}{\sqrt{3}})\\
&\hookrightarrow F_i=\left( u_{Li},~d_{Li},~D_{Li} \right) ~\bigoplus~
Xf_i^c=\left(\nu_{Ri}^{\prime c},~ e_{Ri}^{\prime c},~ e_{Ri}^{c}\right) ~\bigoplus~
u_{Ri}^c~,
\end{align}
where we denote the left-handed (LH) and right-handed (RH) SM fermions respectively as $e_{Li}$, $\nu_{Li}$, $u_{Li}$, $d_{Li}$, and $e_{Ri}$, $u_{Ri}$,  $d_{Ri}$, with $i=1,2,3$ being a generation index. The $U(1)_X$ charges of the fermions are given according to that
the $U(1)_X$ charge operator, denoted as $\hat{Q}_X$, for the $6$ representation 
of the $SU(6)$ group is
\begin{align}
{\hat{Q}_{X}}(6)=\frac{1}{2\sqrt{3}}{\rm diag}[-1,-1,-1,~1,~1,~1]~.
\end{align}
Consequently, the $U(1)_{\text{EM}}$ charge operator can be expressed as 
\begin{equation}
\hat{Q}_{\text{EM}}=\frac{1}{\sqrt{3}}\hat{T}_{8L}+\hat{T}_{3L}+\frac{2}{\sqrt{3}}\hat{Q}_X~,
\end{equation}
where $\hat{T}_{8L}$ and $\hat{T}_{3L}$ are generators of the $SU(3)_L$ gauge group.
Besides, we have fermions, $N_{si}$ and $N^{\prime}_{si}$ ($i=1,2,3$), transforming as singlets under the $SU(3)_C\times SU(3)_L\times U(1)_X$ group.

This 331 model contains 3 scalar triplets from two $\bar{6}$ and one $15$ representations of the $SU(6)$ group, which are
\begin{align}\label{higgs}
15^\prime&\to(1,~\bar{3},~\frac{1}{\sqrt{3}}):
T_u=\frac{1}{\sqrt{2}}
\left( 
\begin{matrix}
v_u+\rho_1+i\sigma_1\\
\sqrt{2}\chi_1^+\\
\sqrt{2}\chi_2^+
\end{matrix}
\right)
,~<T_u>=\frac{1}{\sqrt{2}}
\left( 
\begin{matrix}
v_u\\
0\\
0
\end{matrix}
\right)~,\\
\bar{6}^{\prime\prime}&\to(1,~\bar{3},~\frac{-1}{2\sqrt{3}}):
T_d=\frac{1}{\sqrt{2}}
\left( 
\begin{matrix}
\sqrt{2}\xi_2^-\\
v_d+\rho_2+i\sigma_2\\
\rho_3+i\sigma_3
\end{matrix}
\right)
,~<T_d>=\frac{1}{\sqrt{2}}
\left( 
\begin{matrix}
0\\
v_d\\
0
\end{matrix}
\right)~,\\
\bar{6}^{\prime\prime\prime}&\to (1,~\bar{3},~\frac{-1}{2\sqrt{3}}):
T=\frac{1}{\sqrt{2}}
\left(
\begin{matrix}
\sqrt{2}\xi_1^-\\
\rho_4+i\sigma_4\\
v_t+\rho_5+i\sigma_5
\end{matrix}
\right)
,~<T>=\frac{1}{\sqrt{2}}
\left( 
\begin{matrix}
0\\
0\\
v_t
\end{matrix}
\right)~,
\end{align}
where $v_t$ is the vacuum expectation value (VEV) that breaks the $SU(3)_L\times U(1)_X$ gauge symmetry to the $SU(2)_L\times U(1)_Y$ gauge symmetry \cite{Hoang:1997su,Dong:2013ioa,Pisano:1991ee,Tonasse:1996cx,Nguyen:1998ui}, and $v_u$, $v_d$ are the VEVs which sequently break the $SU(2)_L\times U(1)_Y$ gauge symmetry to the $U(1)_{\text{EM}}$ gauge symmetry. It is known that $\sqrt{v_d^2+v_u^2}\approx 246$ GeV is required to give masses of the SM fermions and we define
\begin{equation}
\tan\theta=\frac{v_u}{v_d}~.
\end{equation}    

The Yukawa terms and Majorana mass terms of this 331 model are
\begin{equation}
\begin{aligned}\label{yukawa}
-\mathcal{L}_{qua}&=y_{ij}^uF_iu_{Rj}^cT_u+y_{ij}^dF_id_{Rj}^cT_d+y_{ij}^DF_iD_{Rj}^cT+H.c,\\
-\mathcal{L}_{lep}&=y_{ij}^\nu f_if_jT_u+y_{ij}^e f_i Xf_j^cT_d+y_{ij}^{L^\prime}f_i^\prime Xf_j^cT+y_{ij}^Nf_i\bar{T}N_{sj}+y_{ij}^{N^\prime}f^{\prime}_i\bar{T}N^\prime_{sj}+H.c, \\
-\mathcal{L}_{neu}^{maj}&=\frac{1}{2}
\left( N_s~~N^\prime_s\right) 
\left\lbrace 
\begin{matrix}
M_s  & M_{ss^\prime}  \\
M_{ss^\prime}^T  & M^\prime_s 
\end{matrix}
\right\rbrace
\left( 
\begin{matrix}
N_s  \\
N^\prime_s  
\end{matrix}
\right) +H.c,
\end{aligned}
\end{equation}
where $M_s$, $M_{s}^\prime$ and $M_{ss^\prime}$ are $3\times 3$ matrix.

The most general Higgs potential in this 331 model is
\begin{align}
V_{\text{Higgs}}=&-m_1^2|T|^2-m_2^2|T_d|^2-m_3^2|T_u|^2+\left(-BT^\dagger T_d+A T_u T_d T+H.c. \right)+V^{(4)}~, \\
V^{(4)}=&l_1\left|T \right|^4+ l_2\left|T_d \right|^4+l_3\left|T_u \right|^4 \nonumber\\ 
&+l_{13}\left|T \right|^2 \left|T_u \right|^2+l_{12}\left|T \right|^2 \left|T_d \right|^2+l_{23}\left|T_u \right|^2 \left|T_d \right|^2+l^{\prime}_{13}\left|T^\dagger T_u \right|^2+l^{\prime}_{12}\left|T^\dagger T_d \right|^2+l^{\prime}_{23}\left|T_u^\dagger T_d \right|^2 \nonumber \\
&+\left(y_1 T^\dagger T_d \left|T\right|^2 +
y_{2} T^\dagger T_d \left|T_d\right|^2+
y_{3} T^\dagger T_d \left|T_u\right|^2+y_{12} T^\dagger T_d T^\dagger T_d+y_{123} T_u^\dagger T_d T_u T ^\dagger+H.c
\right)~,
\end{align}
where $V^{(4)}$ contains all the quartic couplings of the Higgs potential. We can express $B$ and $m^2_i$ ($i=1,2,3$) with other parameters of $V_{\text{Higgs}}$ since $<\frac{\partial V_{\text{Higgs}}}{\partial \rho_j}>=0$ ($j=1,2,...,5$) gives four independent relations.

The non-SM gauge bosons in this 331 model are ${W^{\prime}}^\pm$, $Z^\prime$, and $V/V^*$. Details of the neutral and charged scalar mixings can be found in \cite{Li:2019qxy}. Eigenstates of the mixing of CP-odd scalars $\sigma_{i}$ ($i=1,2,..5$) contain three massless Goldstone bosons $a_1$, $a_2$, $a_3$, and massive
\begin{align}
a_{4}&=-\frac{v_{t}}{\sqrt{v_{d}^{2}+v_{t}^{2}}} \sigma_{3}+\frac{v_{d}}{\sqrt{v_{d}^{2}+v_{t}^{2}}} \sigma_{4}~, \label{a4}\\
a_{5}&=\frac{v_{t} v_{d}}{\sqrt{v_{t}^{2} v_{d}^{2}+v_{t}^{2} v_{u}^{2}+v_{u}^{2} v_{d}^{2}}} \sigma_{1}+\frac{v_{t} v_{u}}{\sqrt{v_{t}^{2} v_{d}^{2}+v_{t}^{2} v_{u}^{2}+v_{u}^{2} v_{d}^{2}}} \sigma_{2}+\frac{v_{u} v_{d}}{\sqrt{v_{t}^{2} v_{d}^{2}+v_{t}^{2} v_{u}^{2}+v_{u}^{2} v_{d}^{2}}} \sigma_{5}~. \label{a5}
\end{align}
The mixing of CP-even scalars $\rho_{i}$ ($i=1,2,..5$) gives four massive eigenstates $h_j$ ($j=2,3,4,5$) and massless
\begin{align}
h_{1}=-\frac{v_{d}}{\sqrt{v_{d}^{2}+v_{t}^{2}}} \rho_{3}+\frac{v_{t}}{\sqrt{v_{d}^{2}+v_{t}^{2}}} \rho_{4}~. \label{h1}
\end{align}

\section{Boundedness from below}\label{BFB}

As there are three scalar triplets in this 331 model, it is necessary to find the criteria for the parameters to ensure a stable scalar potential. The relevant issues can be found in \cite{Ivanov:2018jmz,Maniatis:2006fs,Maniatis:2015gma,Degee:2012sk,Kannike:2012pe,Kannike:2016fmd,Faro:2019vcd,Costantini:2020xrn}. In this Section, we will derive a set of conditions to ensure the BFB of the potential in this 331 model.

The BFB condition is only relevant to the quartic couplings in the scalar potential, namely
$V^{(4)}$
in this model. Without loss of generality, we parameterize a scalar triplet  in the following form \cite{Costantini:2020xrn}
\begin{align}
\Phi_{i}=\sqrt{r_{i}} e^{i \gamma_{i}}\left(\begin{array}{c}{\sin a_{i} \cos b_{i}} \\ {e^{i \beta_{i}} \sin a_{i} \sin b_{i}} \\ {e^{i \alpha_{i}} \cos a_{i}}\end{array}\right), \quad i=1,2,3~,
\end{align}	
where $i$ stands for the label of three scalar triplets in this model.
Each $\Phi_{i}$ contains five real angular parameters, $a_i$, $b_i$, $\alpha_i$, $\beta_i$, $\gamma_i$, and satisfies $\Phi_{i}^\dagger \Phi_{i}=r_i>0$.

The symmetry of $V^{(4)}$ includes a U(3) transformation of $T$, $T_d$, and $T_u$, and a U(1) transformation of $T_u$, which allows us to reduce 10 parameters in the three triplet fields to write
\begin{align}
\frac{T}{\sqrt{r_{1}}}=\left(\begin{array}{l}{0} \\ {0} \\ {1}\end{array}\right),~~~ \frac{T_d}{\sqrt{r_{2}}}=\left(\begin{array}{c}{0} \\ {\sin a_{2}} \\ {e^{i \alpha_{2}} \cos a_{2}}\end{array}\right),~~~ \frac{T_u}{\sqrt{r_{3}}}=\left(\begin{array}{c}{\sin a_{3} \cos b_{3}} \\ {\sin a_{3} \sin b_{3}} \\ {e^{i \alpha_{3}} \cos a_{3}}\end{array}\right)~.
\end{align}	

We assume that all the parameters in $V^{(4)}$ are real for simplicity. By using the reformulated scalar fields, it is direct to get
\begin{align}
V^{(4)}=& l_1 r_1^2+l_2 r_2^2+l_3 r_3^2 +\left(l_{13} +l_{13}^{\prime}\cos^2[a_3]  \right) r_1 r_3  \nonumber\\
&+\left(l_{12} +l_{12}^{\prime}\cos^2[a_2]+2y _{12}  \cos^2[a_2]\cos[2\alpha_2]                \right) r_1 r_2  \nonumber\\
&+2y_1\cos[a_2]\cos[\alpha_2]r_1\sqrt{r_1 r_2}  +2y_2\cos[a_2]\cos[\alpha_2]r_2\sqrt{r_1 r_2}  \nonumber\\
&+\left( l_{23}+l_{23}^{\prime}\left(\cos[a_2]\cos[a_3]\cos[\alpha_2-\alpha_3]+\sin[a_2]\sin[a_3]\sin[b_3] \right)^2\right. +  \nonumber\\
&~~~~\left. l_{23}^{\prime}\left(\cos^2[a_2]\cos^2[a_3]\sin^2[\alpha_2-\alpha_3] \right) \right) r_2 r_3 \nonumber\\
&+2\left(y_{123}\cos[a_3]\left(\cos[a_2]\cos[a_3]\cos[\alpha_2-\alpha_3]+\sin[a_2]\sin[a_3]\sin[b_3]\cos[\alpha_3] \right)+\right. \nonumber\\
&~~~~\left. y_3\cos[a_2]\cos[\alpha_2]  \right) r_3\sqrt{r_1 r_2}~.
\end{align}	

Next, we need to make $V^{(4)}$ independent of the angular parameters by setting $V^{(4)}$ as a ``angular minima'' , for which we define

\begin{align}
k_1=&\frac{\partial V^{(4)}}{\partial \alpha_{3}} \nonumber \\ 
=&r_3\cos[a_3]\big( r_2 l_{23}^\prime  \sin[2a_2]\sin[a_3]\sin[b_3]\sin[\alpha_{2}-\alpha_{3}]  \nonumber\\
&+2y_{123}\sqrt{{r_1}{r_2}}\left(\cos[a_2]\cos[a_3]\sin[\alpha_2-\alpha_{3}]-\sin[a_2]\sin[a_3]\sin[b_3]\sin[\alpha_3] \right) \big)~,\\	
k_2=&\frac{\partial V^{(4)}}{\partial \alpha_{2}} \nonumber \\
=&-\cos[a_2]\big( r_2 r_3 l_{23}^\prime \sin[2a_3]\sin[a_2]\sin[b_3]\sin[\alpha_{2}-\alpha_{3}]+2(y_1r_1+y_2r_2+y_3r_3)\sqrt{r_1r_2}\sin[\alpha_{2}] \nonumber\\
&+4y_{12}r_1r_2\cos[a_2]\sin[2\alpha_2] +2y_{123}r_3\sqrt{r_1r_2}\cos^2[a_3]\sin[\alpha_2-\alpha_3]\big)~,\\
k_3=&\frac{\partial V^{(4)}}{\partial b_{3}} \nonumber\\
=&2r_3\sin[a_2]\sin[a_3]\cos[b_3]\big(y_{123}\sqrt{r_1r_2}\cos[a_3]\cos[\alpha_3] \nonumber\\
&+ r_2l_{23}^\prime \left(\cos[a_2]\cos[a_3]\cos[\alpha_{2}-\alpha_{3}] +\sin[a_2]\sin[a_3]\sin[b_3]\right)  \big)~, \\
k_4=&\frac{\partial V^{(4)}}{\partial a_{3}} \nonumber\\
=&-r_1r_3l_{13}^\prime\sin[2a_3] \nonumber\\
&+2r_3\sqrt{r_1r_2}y_{123}\left(-\sin[2a_3]\cos[a_2]\cos[\alpha_{2}-\alpha_{3}]+\sin[a_2]\sin[b_3]\cos[2a_3]\cos[\alpha_{3}] \right) \nonumber\\
&+r_2r_3l_{23}^\prime\left(-\sin[2a_3]\cos^2[a_2]+\sin[2a_2]\sin[b_3]\cos[2a_3]\cos[\alpha_{2}-\alpha_{3}]+\sin^2[a_2]\sin[2a_3]\sin^2[b_3] \right)~,\\
k_5=&\frac{\partial V^{(4)}}{\partial a_{2}} \nonumber\\
=&-r_1r_2\left(l_{12}^\prime+2y_{12}\cos[2\alpha_{2}] \right) \sin[2a_2] -2(y_1r_1+y_2r_2+y_3r_3)\sqrt{r_1r_2}\sin[a_2]\cos[\alpha_{2}] \nonumber\\
&+2y_{123}r_3\sqrt{r_1r_2}\left(-\sin[a_2]\cos[a_3]\cos[\alpha_{2}-\alpha_{3}]+\sin[a_3]\sin[b_3]\cos[a_2]\cos[\alpha_{3}] \right) \nonumber\\
&+r_2r_3l_{23}^\prime\left(-\sin[2a_2]\cos^2[a_3]  +\sin[b_3]\left(\sin[2a_3]\cos[2a_2]\cos[\alpha_{2}-\alpha_{3}]+\sin[2a_2]\sin^2[a_3]\sin[b_3] \right)  \right)~. 
\end{align}	

The ``angular minima'' requires that all of $k_{i}$ ($i=1,2,\dots,5$) vanish, for which we find four solutions,
\subparagraph*{Solution 1:}
$\sin[a_3]=0$, $\cos[a_2]=0$, $\cos[\alpha_{2}]=0$, $\sin[b_3]=0$, and $
\sin[\alpha_3]=0$, which lead to
\begin{align}
V^{(4)}=&V^{(4)}_1 \nonumber \\
=&l_1 r_1^2+l_2 r_2^2+l_3r_3^2+l_{23}r_2r_3+l_{12}r_1r_2+(l_{13}+l_{13}^\prime)r_1r_3~.
\end{align}

\subparagraph*{Solution 2:}
$\cos[a_2]=0$, $\cos[a_3]=0$, $\cos[\alpha_{2}]=0$, $\sin[2b_3]=0$, and $ \sin[b_3]\cos[\alpha_3]=0$, which lead to
\begin{align}\label{S2}
V^{(4)}=&V^{(4)}_2 \nonumber \\
=&l_1 r_1^2+l_2 r_2^2+l_3r_3^2+l_{13}r_1r_3+l_{12}r_1r_2+(l_{23}+\sin^2[b_3]l_{23}^\prime)r_2r_3 \nonumber\\
\geq & l_1 r_1^2+l_2 r_2^2+l_3r_3^2+l_{13}r_1r_3+l_{12}r_1r_2+(l_{23}+\frac{l_{23}^\prime-|l_{23}^\prime|}{2})r_2r_3~.
\end{align}		
Note that $\sin^2[b_3]$ can be either 0 or 1 as $\sin[2 b_3]=0$ in this solution, which makes that $\sin^2[b_3]l_{23}^\prime \geq \frac{l_{23}^\prime-|l_{23}^\prime|}{2}$ is always true no matter $l_{23}^\prime$ is positive or negative.
\subparagraph*{Solution 3:}
$\sin[a_2]=0$, $\cos[a_3]=0$, $\sin[\alpha_{2}]=0$, and $\sin[b_3]\cos[\alpha_3]=0$, which lead to
\begin{align}
V^{(4)}=&V^{(4)}_3 \nonumber \\
=&l_1 r_1^2+l_2 r_2^2+l_3r_3^2+l_{13}r_1r_3+l_{23}r_2r_3+(l_{12}+l_{12}^\prime+2y_{12})r_1r_2 \nonumber\\
&+2\cos[a_2]\cos[\alpha_2](y_1r_1+y_2r_2+y_3r_3)\sqrt{r_1r_2} \nonumber\\
\geq & l_1 r_1^2+l_2 r_2^2+l_3r_3^2+l_{13}r_1r_3+l_{23}r_2r_3+(l_{12}+l_{12}^\prime+2y_{12})r_1r_2 \nonumber\\
&-(|y_1|r_1+|y_2|r_2+|y_3|r_3)(r_1+r_2)~, 
\end{align}	
where $\cos[a_2]\cos[\alpha_2]=\pm 1$ and $2\sqrt{r_1 r_2}\leq r_1+r_2$ have been utilized.

\subparagraph*{Solution 4:}
$\sin[a_2]=0$, $\sin[a_3]=0$, $\sin[\alpha_{3}]=0$, and $\sin[\alpha_{2}]=0$, which lead to
\begin{align}
V^{(4)}=&V^{(4)}_4 \nonumber \\
=&l_1 r_1^2+l_2 r_2^2+l_3r_3^2+(l_{13}+l_{13}^\prime)r_1r_3+(l_{12}+l_{12}^\prime+2y_{12})r_1r_2+(l_{23}+l_{23}^\prime)r_2r_3 \nonumber \\
&+2\cos[a_2]\cos[\alpha_2](y_1r_1+y_2r_2+y_3r_3)\sqrt{r_1r_2}+2\cos[a_2]\cos[\alpha_{2}-\alpha_{3}]y_{123}r_3\sqrt{r_1r_2} \nonumber \\
\geq &l_1 r_1^2+l_2 r_2^2+l_3r_3^2+(l_{13}+l_{13}^\prime)r_1r_3+(l_{12}+l_{12}^\prime+2y_{12})r_1r_2+(l_{23}+l_{23}^\prime)r_2r_3 \nonumber \\
&-(|y_1|r_1+|y_2|r_2+|y_3|r_3)(r_1+r_2)-|y_{123}|r_3(r_1+r_2)~, 
\end{align}	
where $\cos[a_2]\cos[\alpha_2]=\pm 1$, $\cos[a_2]\cos[\alpha_{2}-\alpha_{3}]=\pm 1$, and $2\sqrt{r_1 r_2}\leq r_1+r_2$ have been utilized.

According to the four solutions given above, we obtain that
\begin{align}\label{MBFB}
V_i^{(4)}\geq (r_1,~r_2,~r_3) M_i^{\text{BFB}} (r_1,~r_2,~r_3)^{\text{T}}~,~~~~~~~i=1,2,3,4.
\end{align}
It is direct to get the expressions of $M_i^{\text{BFB}}$ which we do not give here. The BFB is realized as long as that co-positivity constraints on the four $M_{i}^{\text{BFB}}$ ($i=1,2,3,4$) are satisfied at the same time. 

It is noted that the conditions, derived in this Section, are sufficient but not necessary to ensure the BFB of the potential in our 331 model since the left side in Eq.~\ref{MBFB} could be larger than the right side.

\section{$\text{g}_\mu$-2}\label{gg2}

In this 331 model, new contributions to the muon anomalous moment $a_\mu$ arise from loop diagrams involved the BSM fermions, scalars, and gauge bosons. According to \cite{Sirunyan:2018nnz}, $|v_t|$ is required to be larger than 10 TeV to satisfy the constraint $M_{Z^\prime}>$ 4.5 TeV, which also makes $M_V\approx M_{W^\prime}>$ 3.2 TeV. So the contributions to $a_\mu$ involving the non-SM gauge bosons are negligible due to the heavy masses of these gauge bosons.
We also neglect the contributions given by the non-SM charged scalars and exotic neutrinos since in most viable parameter space they are very heavy.  So, we only focus on the contributions induced by the neutral scalars.

The contributions to $a_\mu$ involving neutral scalars come from the following parts of the Lagrangian of this 331 model, 
\begin{equation}
\mathcal{L}_{lep} \supset -y_{ij}^e f_i Xf_j^c T_d-y_{ij}^{L^\prime}f_i^\prime Xf_j^cT~.
\end{equation}	
Supposing that $y_{ij}^{e}$ and $y_{ij}^{L^\prime}$ are proportional to $\delta_{ij}$ for simplicity, we  define
\begin{align}
& y_{22}^e=y^{\mu}~,\\
&y_{22}^{L^\prime}=y^{\mu^\prime}~.
\end{align}

\begin{figure}[thb]
	\begin{center}
		\includegraphics[width=0.45\textwidth]{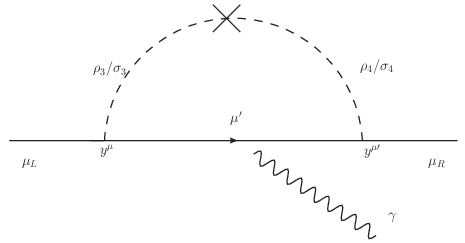}
		\includegraphics[width=0.45\textwidth]{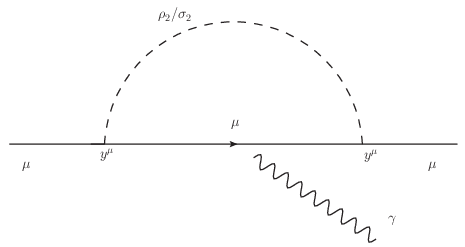}
		\includegraphics[width=0.45\textwidth]{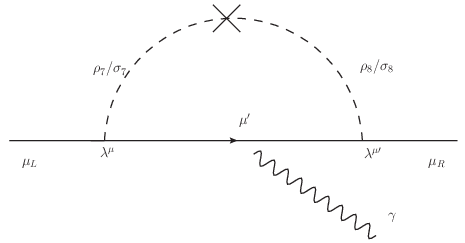}
		\includegraphics[width=0.45\textwidth]{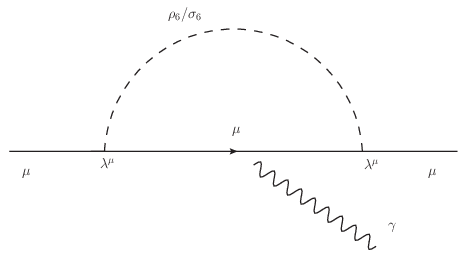}
	\end{center}
	\caption{\label{fig:feynman} Leading contributions to $\Delta a_{\mu}$. }
\end{figure}

Among the contributions involving the non-SM  charged fermions, the leading ones are  chirally-enhanced, which are shown in the upper left panel of FIG.~\ref{fig:feynman}, where the LH and RH $\mu^\prime$ are $e^{\prime}_{L2}$ and $e^{\prime}_{R2}$, respectively. These chirally-enhanced contributions can be expressed as
\begin{align}\label{dmm1}
\Delta a_{\mu 1}=-\frac{m_\mu}{16\pi^2 m_{\mu^\prime}}y^\mu y^{\mu^\prime}\left( \sum_{i=2}^{5} \left(V^h_{3i}V^h_{4i} \right)f_{\text{LR}}^\prime\left( \frac{m_{\mu^\prime}^2}{m_{h_i}^2}\right) -\sum_{j=4}^{5} \left(V^a_{3j}V^a_{4j} \right)f_{\text{LR}}^\prime\left( \frac{m_{\mu^\prime}^2}{m_{a_j}^2}\right) \right) 
\end{align}
where $h_i$ ($a_j$) stands for the eigenstate of the CP-even (CP-odd) scalar mixing, $V^h$ ($V^a$) is the corresponding matrix for the mixing, and
\begin{equation}
f_{\text{LR}}^\prime(x)=x\frac{3-4x+x^2+2\log x}{2(x-1)^3}~.
\end{equation}
Note that $h_1$ and $a_{j}$ ($j=1,2,3$) are not included in the summations in Eq.~\ref{dmm1} because they are Goldstone bosons.

The $f_{\text{LR}}^\prime(x)$ increases with $x$ and satisfies
\begin{align}\label{ff}
f_{\text{LR}}^\prime(x)< \frac{1}{2}~.
\end{align}
It is know that $y^{\mu^\prime}$ is related to the mass of $\mu^\prime$ by
\begin{align}
m_{\mu^\prime}=y^{\mu^\prime}\frac{v_t}{\sqrt{2}}~.
\end{align}
According to Eq.~\ref{a4}, Eq.~\ref{a5}, and Eq.~\ref{h1}, we obtain that
\begin{align}
&\left|V^h_{3i}V^h_{4i}\right|=\left( {V^h_{3i}}\right) ^2\left| \frac{v_d}{v_t}\right|<\left| \frac{v_d}{v_t}\right|~,~~~~~~~~~~~~i=2,3,4,5,\\ 
&\left|V^a_{34}V^a_{44}\right|=\frac{|v_tv_d|}{v_t^2+v_d^2}<\left| \frac{v_d}{v_t}\right|~,\\
&V^a_{35}V^a_{45}=0~. \label{aa5}
\end{align}
Considering Eq.~\ref{ff} to Eq.~\ref{aa5}, we can estimate that
\begin{align}
\left| \Delta a_{\mu 1}\right| \sim \frac{1}{16\pi^2} \left(\frac{m_{\mu}}{v_t} \right)^2\sim 10^{-12}~,
\end{align}
where $|v_t|>10$ TeV has been used.

The contribution involving muon and the non-SM scalar is shown in the upper right panel of FIG.~\ref{fig:feynman}, which can be expressed as
\begin{align}\label{dmm2}
\Delta a_{\mu 2}=&\frac{{y^\mu}^2}{16\pi^2}\left( \sum_{i=3}^{5} \left(V^h_{2i} \right)^2\left( f_{\text{LR}}^\prime\left( \frac{m_{\mu}^2}{m_{h_i}^2}\right)+f_{\text{LL}}^\prime\left( \frac{m_{\mu}^2}{m_{h_i}^2}\right)\right)\right.  \nonumber \\
&\left. +\sum_{j=4}^{5} \left(V^a_{2j}\right)^2\left(- f_{\text{LR}}^\prime\left( \frac{m_{\mu}^2}{m_{a_j}^2}\right)+f_{\text{LL}}^\prime\left( \frac{m_{\mu}^2}{m_{a_j}^2}\right)\right)  \right)~,
\end{align}
where $h_2$ (the Higgs boson) is not included because the relevant contribution is not BSM, and
\begin{equation}
f_{\text{LL}}^\prime(x)=x\frac{2+3x-6x^2+x^3+6x\log x}{6(1-x)^4}~.
\end{equation}
Since $f_{\text{LL}}^\prime(x)\ll f_{\text{LR}}^\prime(x)$ when $x=\frac{m_\mu^2}{m_{h_i/a_j}^2}\ll 1$, the magnitude of $\Delta a_{\mu 2}$ can be estimated as
\begin{align}
\left| \Delta a_{\mu 2}\right| \sim &\frac{{y^\mu}^2}{16\pi^2}f_{\text{LR}}^\prime\left( \frac{m_{\mu}^2}{m_{h_i/a_j}^2}\right)~,
\end{align}
which will be about $\sim$ $0.4\times10^{-13}$ if we choose that $\tan\theta=6$ and $m_{h_i/a_j}=600$ GeV.

We conclude that it is impossible to account for the muon anomalous moment with the scalars now available. To account for the muon anomalous moment in this 331 model, we introduce another two 
scalar triplets, $T^\prime$ and $T_d^\prime$, which are in the same representation with $T$ ($T_d$) but have no VEVs
\begin{align}
&T_d^\prime=\frac{1}{\sqrt{2}}
\left( 
\begin{matrix}
\sqrt{2}\xi_3^-\\
\rho_6+i\sigma_6\\
\rho_7+i\sigma_7
\end{matrix}
\right)~,\\
&T^\prime=\frac{1}{\sqrt{2}}
\left(
\begin{matrix}
\sqrt{2}\xi_4^-\\
\rho_8+i\sigma_8\\
\rho_9+i\sigma_9
\end{matrix}
\right)~.
\end{align}
Very similar to~\cite{Calibbi:2019bay}, the following terms can be added in the Lagrangian of this 331 model
\begin{align}
\Delta\mathcal{L}&=\left(- {\lambda_{ij}^e f_i Xf_j^c T_d^\prime}-{ \lambda_{ij}^{L^\prime}f_i^\prime Xf_j^cT^\prime}+A^\prime T_uT_d^\prime T^\prime+H.c.\right) -m_4^2 \left| T^\prime\right|^2-m_5^2 \left| T_d^\prime\right|^2~.
\end{align}
It should be emphasized that we do not add all the gauge invariant terms involving $T^\prime$ and $T_d^\prime$ for simplicity.

New mixing of neutral scalars arises. The mass matrix and the mixing matrix satisfy
\begin{align}
{U^{1,2}}^{\dagger}\left(\begin{array}{cc}
m_{4}^{2} & \pm A^{\prime} v_u / \sqrt{2} \\
\pm A^{\prime} v_u / \sqrt{2} & m_{5}^{2}
\end{array}\right) U^{1,2}=\left(\begin{array}{cc}
M_{{1}}^{2} & \\
& M_{{2}}^{2}
\end{array}\right)~,
\end{align}
where $+A^{\prime} v_u / \sqrt{2}$ in the mass  matrix in company with $U^1$ is for the mixing of $\rho_7$ and $\rho_8$ (also for the mixing of $\sigma_6$ and $\sigma_9$), and $-A^{\prime} v_u / \sqrt{2}$ in the mass matrix together with $U^2$ is for the mixing of $\rho_6$ and $\rho_9$ (also for the mixing of $\sigma_7$ and $\sigma_8$). It is direct to obtain the physical masses and mixing, which are
\begin{align}
M_{{1,2}}^{2}=\left(m_{4}^{2}+m_{5}^{2} \mp \Delta M^{2}\right) / 2, \quad \Delta M^{2} \equiv \sqrt{\left(m_{4}^{2}-m_{5}^{2}\right)^{2}+2 {A^{\prime}}^{2} v_u^{2}}~,
\end{align}
\begin{align}
U^{1,2}=\left(\begin{array}{cc}
\frac{\pm \sqrt{2} A^{\prime} v_u}{\sqrt{\left(m_{5}^{2}-m_{4}^{2}-\Delta M^{2}\right)^{2}+2 {A^{\prime}}^{2} v_u^{2}}} & -\frac{m_{5}^{2}-m_{4}^{2}-\Delta M^{2}}{\sqrt{\left(m_{5}^{2}-m_{4}^{2}-\Delta M^{2}\right)^{2}+2 {A^{\prime}}^{2} v^{2}}} \\
\frac{m_{5}^{2}-m_{4}^{2}-\Delta M^{2}}{\sqrt{\left(m_{5}^{2}-m_{4}^{2}-\Delta M^{2}\right)^{2}+2 {A^{\prime}}^{2} v_u^{2}}} & \frac{\pm \sqrt{2} A^{\prime} v_u}{\sqrt{\left(m_{5}^{2}-m_{4}^{2}-\Delta M^{2}\right)^{2}+2 {A^{\prime}}^{2} v_u^{2}}}
\end{array}\right)~.
\end{align}

Supposing that $\lambda_{ij}^{e}$ and $\lambda_{ij}^{L^\prime}$ are proportional to $\delta_{ij}$ for simplicity, we  define
\begin{align}
& \lambda_{22}^e=\lambda^{\mu}~,\\
&\lambda_{22}^{L^\prime}=\lambda^{\mu^\prime}~.
\end{align}

Similarly, the leading contributions involving $\mu^\prime$ and neutral scalars from $T^\prime$ and $T_d^\prime$ are the chirally-enhanced ones shown in the lower left panel of FIG.~\ref{fig:feynman}, which can be expressed as
\begin{align}\label{d1m}
\Delta a_{\mu 3}&=-\frac{m_\mu}{16\pi^2 m_{\mu^\prime}}\lambda^\mu \lambda^{\mu^\prime}\sum_{i=1}^{2}\left(  \left(U^1_{1i} U^1_{2i} \right)f_{\text{LR}}^\prime\left( \frac{m_{\mu^\prime}^2}{M_{i}^2}\right) - \left(U^2_{1i} U^2_{2i} \right)f_{\text{LR}}^\prime\left( \frac{m_{\mu^\prime}^2}{M_{i}^2}\right) \right) \nonumber\\
&=-\frac{m_\mu}{8\pi^2 m_{\mu^\prime}}\lambda^\mu \lambda^{\mu^\prime}\left( \sum_{i=1}^{2} \left(U^1_{1i} U^1_{2i} \right)f_{\text{LR}}^\prime\left( \frac{m_{\mu^\prime}^2}{M_{i}^2}\right) \right)~,
\end{align}
where $U^1_{1i} U^1_{2i}=-U^2_{1i} U^2_{2i}$ ($i=1,2$) is used.

The contribution involving muon and neutral scalars from $T_d^\prime$ is shown in the lower right panel of FIG.~\ref{fig:feynman}, which can be expressed as
\begin{align}\label{d2m}
\Delta a_{\mu 4}=&\frac{{\lambda^\mu}^2}{16\pi^2} \sum_{i=1}^{2}\left( \left(U^2_{1i} \right)^2\left( f_{\text{LR}}^\prime\left( \frac{m_{\mu}^2}{M_{i}^2}\right)+f_{\text{LL}}^\prime\left( \frac{m_{\mu}^2}{M_{i}^2}\right)\right) + \left(U^1_{1i}\right)^2\left(- f_{\text{LR}}^\prime\left( \frac{m_{\mu}^2}{M_{i}^2}\right)+f_{\text{LL}}^\prime\left( \frac{m_{\mu}^2}{M_{i}^2}\right)\right)  \right) \nonumber \\
=&\frac{{\lambda^\mu}^2}{8\pi^2}\left( \sum_{i=1}^{2} \left(U^1_{1i} \right)^2 f_{\text{LL}}^\prime\left( \frac{m_{\mu}^2}{M_{i}^2}\right)\right)~,
\end{align}
where $\left( U^1_{1i}\right)^2=\left( U^2_{1i}\right)^2$ ($i=1,2$) is used.

Based on the above discussions, we only need to consider contributions $\Delta a_{\mu 3}$ and $\Delta a_{\mu 4}$, and thus have
\begin{align}
\Delta a_\mu \approx \Delta a_{\mu 3}+\Delta a_{\mu 4}~.
\end{align}

\begin{figure}[thb]
	\begin{center}
		\includegraphics[height=0.35\textwidth]{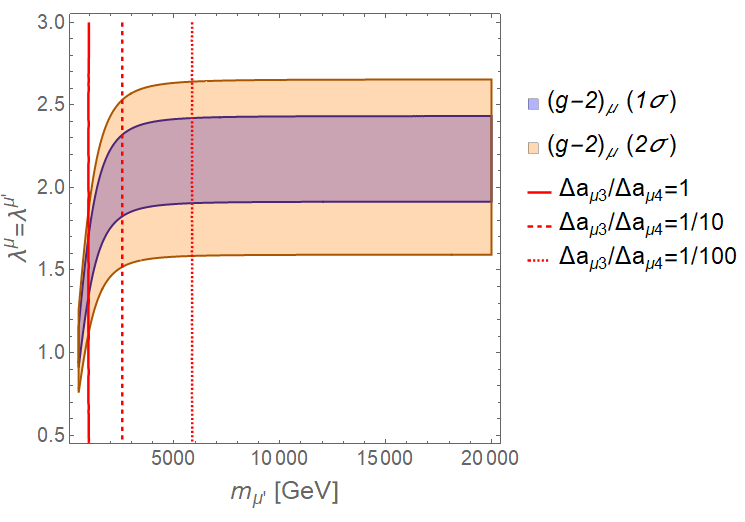}
		\includegraphics[height=0.35\textwidth]{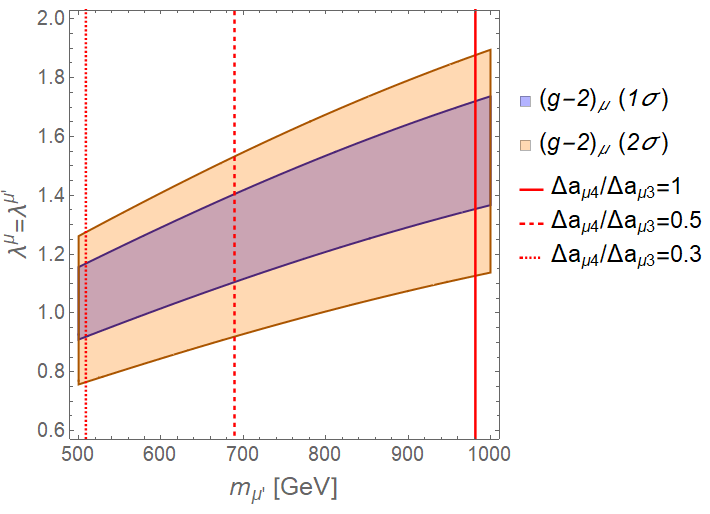}\\
		\includegraphics[height=0.35\textwidth]{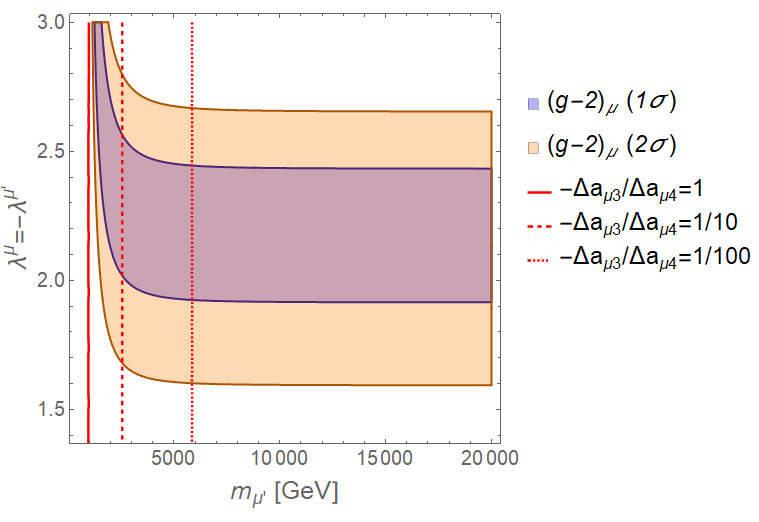}
	\end{center}
	\caption{\label{fig:g-2mu} Leading contributions to $\Delta a_\mu$ as a function of $m_{\mu^\prime}$ and $\lambda^\mu$. Here $\lambda^{\mu}=\pm\lambda^{\mu^\prime}$ and other parameters are set to: $m_4=300$ GeV, $m_5=600$ GeV, $A^\prime=10$ GeV, $\tan\theta=6$. }
\end{figure}

To study the dependences of $\Delta a_{\mu}$ on $\lambda^\mu$, ${\lambda^\mu}^\prime$, and $m_{\mu^\prime}$, we set other parameters to that $m_4=300$ GeV, $m_5=600$ GeV, $A^\prime=10$ GeV, and $\tan\theta=6$. In FIG.~\ref{fig:g-2mu}, points in the orange (blue) region on the $m_{\mu^\prime}-\lambda^\mu$ plane give values of $\Delta a_\mu$ within $2\sigma$ ($1\sigma$) deviation from the measured value. 

In the upper two panels of FIG.~\ref{fig:g-2mu} where $\lambda^\mu={\lambda^\mu}^\prime$, it is shown that $a_{\mu 4}/a_{\mu 3}$ increases with $m_{\mu^\prime}$ and $\Delta a_{\mu 4}$ dominates ($a_{\mu 4}/a_{\mu 3}>1$) when $m_{\mu^\prime}$ is larger than 1 TeV. When $m_{\mu^\prime}$ is larger than 5.9 TeV, $\Delta a_{\mu 4}$ contributes more than 99\% of $\Delta a_{\mu}$ and both the $1\sigma$ and $2\sigma$ regions have no dependence on $m_{\mu^\prime}$ since $\Delta a_{\mu 4}$ is independent of $m_{\mu^\prime}$. In the $m_{\mu^\prime}>$5.9 TeV region, $\lambda^\mu={\lambda^\mu}^\prime$  within the $1\sigma$ and $2\sigma$ regions needs to be in the ranges of ($1.92,2.43$) and ($1.60,2.65$), respectively. $\Delta a_{\mu 3}$ dominates ($a_{\mu 4}/a_{\mu 3}<1$) when $m_{\mu^\prime}$ is less than 1 TeV and the allowed $\lambda^\mu={\lambda^\mu}^\prime$ within the $1\sigma$ and $2\sigma$ regions increases with $m_{\mu^\prime}$.

In the lower panel of FIG.~\ref{fig:g-2mu} where $\lambda^\mu=-{\lambda^\mu}^\prime$, $\Delta a_{\mu4}$ is positive while
$\Delta a_{\mu3}$ is negative. Both the $1\sigma$ and $2\sigma$ regions need to ensure that $m_{\mu^\prime}>1$ TeV to make $\Delta a_{\mu}>0$. In the $1\sigma$ region, $\lambda^\mu=-{\lambda^\mu}^\prime$ decreases with $m_{\mu^\prime}$ since the negative $\Delta a_{\mu3}$ approaches zero when $m_{\mu^\prime}$ increases. Similarly, $\Delta a_{\mu 4}$ contributes almost the whole $\Delta a_{\mu}$ ($-\Delta a_{\mu 3}/\Delta a_{\mu 4}<0.01$) when $m_{\mu^\prime}$ is larger than 5.9 TeV and the $1\sigma$ and $2\sigma$ regions are approximately independent of $m_{\mu^\prime}$.

In principle, we can add extra terms like $\left( - {\lambda_{ij}^e f_i Xf_j^c T}-{ \lambda_{ij}^{L^\prime}f_i^\prime Xf_j^c T_d}\right)$ to account for the $\Delta a_{\mu}$ without introducing $T^\prime$ and $T_d^\prime$. However, the VEV of $T$ ($T_d$) will lead to mixing between $\mu_L$ and $\mu_L^\prime$ ($\mu_R$ and $\mu_R^\prime$), which can bring about serious fine tuning problem. So we introduce $T^\prime$ and $T_d^\prime$ with no VEVs to account for the $\Delta a_{\mu}$ in this Section.

\section{$\text{g}_e$-2}\label{gge}

Much like the discussions for $\text{g}_\mu$-2, the non-SM particles in this 331 model (with $T^\prime$ and $T_d^\prime$ included) also induce new contributions to the electron anomalous moment $a_e$. Similarly, we define
\begin{align}
& y_{11}^e=y^{e}~,\\
& \lambda_{11}^e=\lambda^{e}~,\\
&y_{11}^{L^\prime}=y^{e^\prime}~,\\
&\lambda_{11}^{L^\prime}=\lambda^{e^\prime}~.
\end{align}
There is no doubt that the leading contributions to $a_e$ can be expressed as
\begin{align}
&\Delta a_{e}\approx \Delta a_{e3}+\Delta a_{e4}~,\\
&\Delta a_{e3}=-\frac{m_e}{8\pi^2 m_{e^\prime}}\lambda^e \lambda^{e^\prime}\left( \sum_{i=1}^{2} \left(U^1_{1i} U^1_{2i} \right)f_{\text{LR}}^\prime\left( \frac{m_{e^\prime}^2}{M_{i}^2}\right) \right)~,\\
&\Delta a_{e4}=\frac{{\lambda^e}^2}{8\pi^2}\left( \sum_{i=1}^{2} \left(U^1_{1i} \right)^2 f_{\text{LL}}^\prime\left( \frac{m_{e}^2}{M_{i}^2}\right)\right)~.
\end{align}

\begin{figure}[thb]
	\begin{center}
		\includegraphics[height=0.35\textwidth]{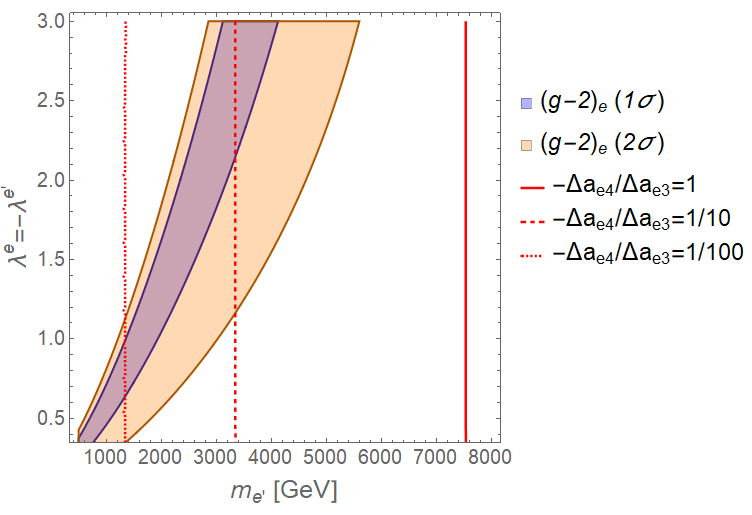}
		\includegraphics[height=0.35\textwidth]{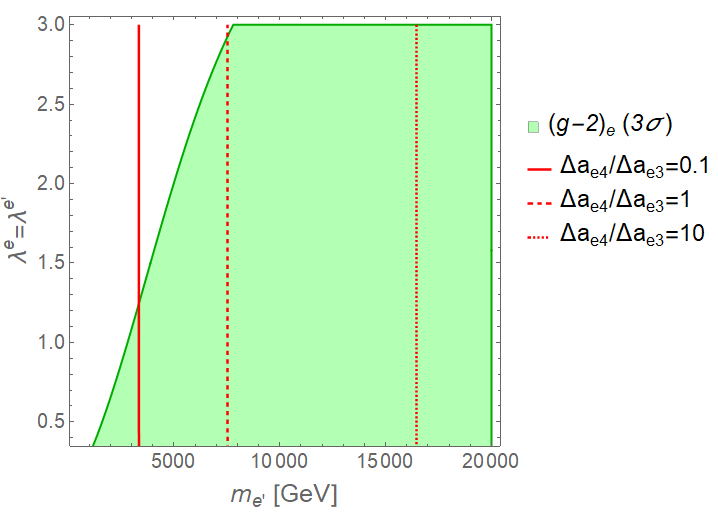}
	\end{center}
	\caption{\label{fig:g-2e}Leading contributions to $\Delta a_e$ as a function of $m_{e^\prime}$ and $\lambda^e$. Here $\lambda^{e}=\pm \lambda^{e^\prime}$ and other parameters are set as in FIG~\ref{fig:g-2mu}.}
\end{figure}

We focus on the dependence of $\Delta a_{e}$ on $\lambda^e$, ${\lambda^e}^\prime$, and $m_{e^\prime}$ by setting other parameters as in FIG~\ref{fig:g-2mu}. Points in the blue, orange, and green regions on the $m_{e^\prime}-\lambda^e$ plane in FIG.~\ref{fig:g-2e} give values of $\Delta a_e$ within $1\sigma$, $2\sigma$ and $3\sigma$ deviations from the measured value, respectively.

In the left panel of FIG~\ref{fig:g-2mu} where $\lambda^e=-{\lambda^e}^\prime$, $\Delta a_{e3}$ is negative while $\Delta a_{e4}$ is positive. The $1\sigma$ and $2\sigma$ regions, which need to give negative value of $\Delta a_{e}$, must be at the left side of the line where $-\Delta a_{e4}/\Delta a_{e3}=1$. So the allowed $m_{e^\prime}$ within the $1\sigma$ ($2\sigma$) region  has a largest value, which is 4.1 TeV (5.6 TeV) when $\lambda^e=-{\lambda^e}^\prime=3$. 

Because $\Delta a_{e3}$ and $\Delta a_{e4}$ are both positive in the right panel of FIG~\ref{fig:g-2mu} where $\lambda^e={\lambda^e}^\prime$, the $1\sigma$ and $2\sigma$ regions disappear. Since $\Delta a_{e4}=1.1\times 10^{-13}$ when $\lambda^e =3$, the region where $\lambda^e <3$ is always within the $3\sigma$ region as long as $\Delta a_{e4}$ dominates ($|\Delta a_{e4}/\Delta a_{e3}|>1$). 

\section{Conclusion}\label{conclu}

We derive a set of sufficient conditions to ensure the boundedness from below of the potential in the 331 model proposed in \cite{Li:2019qxy} which has three scalar triplets. Since the quartic couplings $V^{(4)}$ are more complex than those in \cite{Costantini:2020xrn}, inequalities have to be used during the derivations which makes that the conditions obtained are sufficient but not necessary to ensure the BFB of the potential. 

We focus on the BSM contributions to $\Delta a_\mu$ and $\Delta a_e$ involving neutral scalars and charged fermions. The analysis shows that the contributions induced by the neutral scalars from $T$ and $T_d$ are too small to account for the muon and electron anomalous moments unless there are some serious fine tuning problems. So another two triplets $T^\prime$ and $T_d^\prime$ with non VEVs are introduced to provide new couplings to fermions. With the neutral scalars from $T^\prime$ and $T_d^\prime$, the leading contributions to $\Delta a_\mu$ ($\Delta a_e$) involve both the BSM $\mu^\prime$ ($e^\prime$) and $\mu$ ($e$) from the SM. The chirally-enhanced contributions dominate the contributions from $\mu^\prime$ ($e^\prime$), whose magnitudes decrease with $m_{\mu^\prime}$ ($m_{e^\prime}$), while the contributions from $\mu$ ($e$) are independent of $m_{\mu^\prime}$ ($m_{e^\prime}$). The former dominates over the latter when $m_{\mu^\prime}<1$ TeV ($m_{e^\prime}<7.5$ TeV) if the other parameters are set as the Section~\ref{gg2}.

\newpage
\section*{Acknowledgement}
The research was supported by the Projects 11875062 and 11947302 supported by the National Science Foundation of China, and by the Key Research Program of Frontier Science, CAS.

\phantomsection
\addcontentsline{toc}{section}{References}
\bibliographystyle{jhep}
\bibliography{331g2}

\end{document}